\documentclass[sigconf,natbib=true,anonymous=false]{acmart}
\pdfoutput=1

\AtBeginDocument{%
  \providecommand\BibTeX{{%
    \normalfont B\kern-0.5em{\scshape i\kern-0.25em b}\kern-0.8em\TeX}}}

\setcopyright{acmcopyright}
\copyrightyear{2018}
\acmYear{2018}
\acmDOI{10.1145/1122445.1122456}

\acmConference[Woodstock '18]{Woodstock '18: ACM Symposium on Neural
  Gaze Detection}{June 03--05, 2018}{Woodstock, NY}
\acmBooktitle{Woodstock '18: ACM Symposium on Neural Gaze Detection,
  June 03--05, 2018, Woodstock, NY}
\acmPrice{15.00}
\acmISBN{978-1-4503-XXXX-X/18/06}

\begin{document}

\title{SEMINAR: Search Enhanced Multi-modal Interest Network and Approximate Retrieval for Lifelong Sequential Recommendation}

\author{Kaiming Shen}
\affiliation{%
  \institution{Ant Group}
    \city{Beijing}
  \country{China}
}
\email{kaiming.skm@antgroup.com}

\author{Xichen Ding}
\affiliation{%
  \institution{Ant Group}
    \city{Beijing}
  \country{China}
}
\email{xichen.dxc@antgroup.com}

\author{Zixiang Zheng}
\affiliation{%
  \institution{Ant Group}
    \city{Beijing}
  \country{China}
}
\email{zhengzixiang.zzx@antgroup.com}

\author{Yuqi Gong}
\affiliation{%
  \institution{Ant Group}
  \city{Beijing}
  \country{China}
  }
\email{gongyuqi.gyq@antgroup.com}

\author{Qianqian Li}
\affiliation{%
  \institution{Ant Group}
    \city{Beijing}
  \country{China}
}
\email{zixi.lqq@antgroup.com}

\author{Zhongyi Liu}
\affiliation{%
  \institution{Ant Group}
      \city{Hangzhou}
  \country{China}
}
\email{zhongyi.lzy@antgroup.com}

\author{Guannan Zhang}
\affiliation{%
  \institution{Ant Group}
    \city{Hangzhou}
  \country{China}
}
\email{zgn138592@antgroup.com}

\renewcommand{\shortauthors}{First Author, et al.}


\begin{abstract}

The modeling of users' behaviors is crucial in modern recommendation systems. A lot of research focuses on modeling users' lifelong sequences, which can be extremely long and sometimes exceed thousands of items. These models use the target item to search for the most relevant items from the historical sequence. However, training lifelong sequences in click through rate (CTR) prediction or personalized search ranking (PSR) is extremely difficult due to the insufficient learning problem of ID embedding, especially when the IDs in the lifelong sequence features do not exist in the samples of training dataset. Additionally, existing target attention mechanisms struggle to learn the multi-modal representations of items in the sequence well. The distribution of multi-modal embedding (text, image and attributes) output of user's interacted items are not properly aligned and there exist divergence across modalities. We also observe that users' search query sequences and item browsing sequences can fully depict users' intents and benefit from each other. To address these challenges, we propose a unified lifelong multi-modal sequence model called SEMINAR-Search Enhanced Multi-Modal Interest Network and Approximate Retrieval. Specifically, a network called Pretraining Search Unit (PSU) learns the lifelong sequences of multi-modal query-item pairs in a pretraining-finetuning manner with multiple objectives: multi-modal alignment, next query-item pair prediction, query-item relevance prediction, etc. After pretraining, the downstream model, which shares the same target attention structure with PSU, restores the pretrained embedding as initialization and finetunes the network. To accelerate the online retrieval speed of multi-modal embedding, we propose a multi-modal codebook-based product quantization strategy to approximate the exact attention calculation and significantly reduce the time complexity.

\end{abstract}

\ccsdesc[500]{Information systems~Recommender systems}
\ccsdesc[300]{Information systems~Data mining}

\keywords{lifelong sequence modeling, multi-modal retrieval}



\begin{abstract}

The modeling of users' behaviors is crucial in modern recommendation systems. A lot of research focuses on modeling users' lifelong sequences, which can be extremely long and sometimes exceed thousands of items. These models use the target item to search for the most relevant items from the historical sequence. However, training lifelong sequences in click through rate (CTR) prediction or personalized search ranking (PSR) is extremely difficult due to the insufficient learning problem of ID embedding, especially when the IDs in the lifelong sequence features do not exist in the samples of training dataset. Additionally, existing target attention mechanisms struggle to learn the multi-modal representations of items in the sequence well. The distribution of multi-modal embedding (text, image and attributes) output of user's interacted items are not properly aligned and there exist divergence across modalities. We also observe that users' search query sequences and item browsing sequences can fully depict users' intents and benefit from each other. To address these challenges, we propose a unified lifelong multi-modal sequence model called SEMINAR-Search Enhanced Multi-Modal Interest Network and Approximate Retrieval. Specifically, a network called Pretraining Search Unit (PSU) learns the lifelong sequences of multi-modal query-item pairs in a pretraining-finetuning manner with multiple objectives: multi-modal alignment, next query-item pair prediction, query-item relevance prediction, etc. After pretraining, the downstream model, which shares the same target attention structure with PSU, restores the pretrained embedding as initialization and finetunes the network. To accelerate the online retrieval speed of multi-modal embedding, we propose a multi-modal codebook-based product quantization strategy to approximate the exact attention calculation and significantly reduce the time complexity.

\end{abstract}
\maketitle

\vspace{-1.0em}
\section{Introduction}

Users' behavior modeling is extremely important in modern commercial recommendation systems, including online e-commerce platforms such as Amazon, Taobao, Alipay, and content platforms such as YouTube, TikTok, etc. As users spend more time on online shopping and watching short videos, the length of users' historical behaviors has grown dramatically from a few hundreds ($10^{2}$) to more than ten-thousands ($10^{4}$) in recent years. A lot of recent research focuses on modeling users' lifelong behaviors, such as Efficient Target Attention (ETA) \cite{chen2022efficient}, Two-Stage Interest Network (TWIN) \cite{10.1145/3580305.3599922}, Query-Dominant Interest Network (QIN) \cite{10.1145/3583780.3615022}, etc. These models follow a cascading two-stage paradigm, which first uses the target item or target search query as a trigger to retrieve the top-K relevant behaviors from historical behaviors. In the second stage, it uses target attention to encode the selected behaviors as users' interest representation. This paradigm is widely adopted in many search and recommendation tasks, such as click
through rate (CTR) prediction and personalized search ranking. The item representations in the sequence are computed using both the item ID feature and more generic attributes' features. One easily neglected problem in existing lifelong behavior modeling is the insufficient learning problem of ID features in the lifelong sequence, such as historical item ID, author ID, etc. Many historical items in the lifelong sequence can't be found in the current training dataset, which is collected from the most recent logs of exposures and clicks. These low frequency ID embeddings can't be learned well by the limited dataset after being randomly initialized, which will harm the accuracy of target attention calculation.

The second problem in existing lifelong sequence modeling is that it can't handle multi-modal features of items in the sequence well, such as text and image features. The norm values of vectors from different modalities vary if the modalities are not properly aligned in the same embedding space. Existing target item attention calculation uses the inner product of query and keys, which may be dominated by modality vectors with large norm values. For example, the target item will only retrieve the top-K visually relevant but semantically very different items from historical behaviors, which will deteriorate the online performance of recommendation.

To tackle these problems, we propose a new model called Search Enhanced Multi-Modal Interest Network and Approximate Retrieval (SEMINAR) to model users' lifelong historical multi-modal behaviors. The users' historical behaviors include heterogeneous behaviors of both the sequence of browsing item and the sequence of search query. We align users' search query sequence with browsing item sequence together as a unified sequence of query-item pairs, which can be retrieved flexibly by target item or target search query in both the CTR prediction task and Personalized Search Ranking (PSR) task. SEMINAR proposes a Pretraining Search Unit (PSU) network to learn the lifelong behavior sequence of historical multi-modal query-item pairs. It introduces multiple pretraining tasks designed to solve the insufficient learning issue of historical ID features and the multi-modal alignment. In downstream tasks, the target attention module restores the learned item representations from PSU, using the pretrained ID embedding as initialization, and applies a projection weight matrix to get the transformed representation of the behavior sequence. 
During online serving, calculating exact attention using the inner product of multi-modal vectors in the lifelong sequence has the time complexity of $O(L \times M \times d)$, which is time consuming. $L$ denotes the sequence length, $M$ denotes the number of modalities and $d$ denotes the embedding dimension.
Different from existing approximate retrieval methods, such as Locally Sensitive Hash (LSH) and Hierarchical Navigable Small World (HNSW), we exploit an approximate strategy of Product Quantization in a multi-modal setting and express the multi-modal item representations as discrete integer codes using the quantization codebooks, and sum the inner product of centroids in sub-vectors to approximate the exact attention calculation. During online serving, the attention calculation is equivalent to pre-computed distance table lookup and summation operations, which can be conducted efficiently.

In summary, the main contributions of our work are as follows:
\begin{itemize}
\item{We identify the insufficient learning problem of ID features in lifelong behavior modeling and observe that target attention calculation is dominated by multi-modal features with large norm values. And we novelly propose SEMINAR framework, which includes the Pretraining Search Unit to effectively alleviate the insufficient learning problems of ID embedding and multi-modal alignment.}

\item{We exploit a product quantization approximation strategy in a multi-modal setting, which can reduce the time complexity during online serving of retrieval using target query item pair from historical behaviors.}

\item{We conduct extensive experiments on real-world datasets to demonstrate the effectiveness of our proposed model. And we also released the code of SEMINAR in this repository \footnote{\url{https://github.com/paper-submission-coder/SEMINAR}} to encourage further research.}
\end{itemize}

\section{Related Work}

\begin{figure*}
  \includegraphics[height=3.2in, width=6.4in]{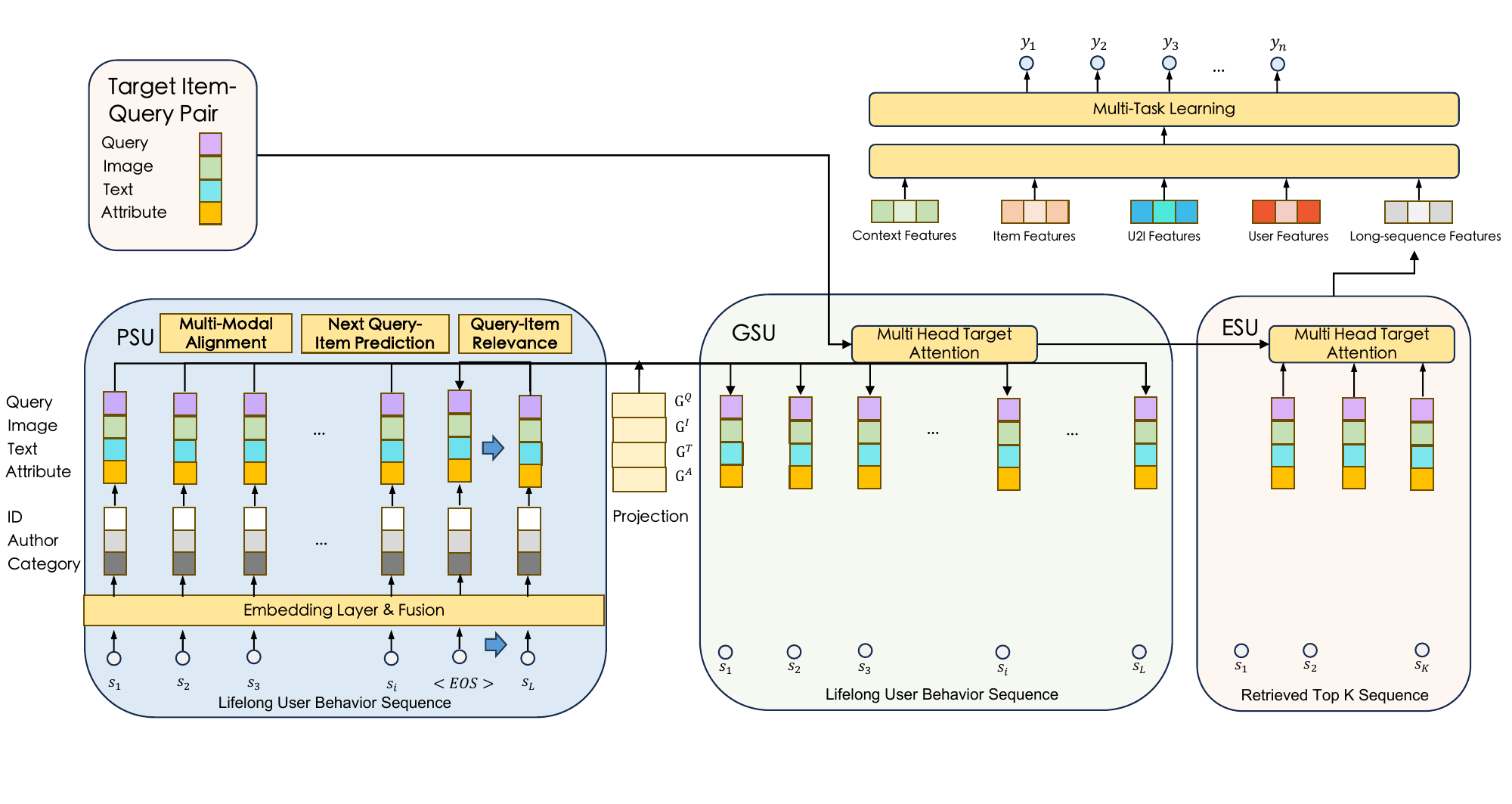}
  \caption{Illustration of SEMINAR Model Architecture. $S_{i}$ denotes the i-th behavior of query and item pair in the lifelong sequence. Each behavior has multiple channels of query and multi-modal features of text, image and attributes. PSU denotes the pretraining search unit. GSU and ESU denote the general and exact search unit respectively as the two stage paradigm. }
  \label{fig:seminar_model_architecture}
\end{figure*}

\subsection{Long-Term Lifelong User Behavior Modeling}
Long-term lifelong user behavior modeling has attracted much research attention in recent years. Typical works include SIM \cite{10.1145/3340531.3412744}, ETA \cite{chen2022efficient}, TWIN \cite{10.1145/3580305.3599922}, QIN \cite{10.1145/3583780.3615022}, etc. SIM \cite{10.1145/3340531.3412744} introduces the General Search Unit to retrieve the top-K most relevant items from historical behaviors using the target item as a trigger, and the Exact Search Unit (ESU) to calculate the multi-head target attention (MHTA). ETA \cite{chen2022efficient} uses a set of hash functions to express the item representation as binary hash embedding and calculates the Hamming distance to approximate the inner product calculation. TWIN \cite{10.1145/3580305.3599922} introduces the CP-GSU as a consistency-preserved lifelong user behavior modeling module to increase the relevance calculation consistency between the two cascading stages. QIN \cite{10.1145/3583780.3615022} uses the search query as a trigger to retrieve the most relevant items from the historical behaviors in the first stage of the cascading models in Personalized Search Ranking. Different from the existing work, we propose the pretraining search unit (PSU) to alleviate the insufficient learning problem of ID features and multi-modal alignment in attention calculation. Furthermore, there is an increasing trend of modeling search and recommendation tasks jointly in a unified framework, such as USER \cite{10.1145/3459637.3482489}, SESRec \cite{10.1145/3539618.3591786}, S\&R Foundation \cite{10.1145/3583780.3614657}, etc. To model the lifelong behaviors, we align the historical search query sequence and browsing item sequence as a unified sequence of query-item pairs, which can be applied to both CTR prediction in recommendation and personalized search ranking.

\subsection{Multi-Modal Alignment in Recommendation and Item Quantization}

Multi-modal alignment is a prevalent topic, which aligns the multi-modal features such as text and image in a unified embedding space in a contrastive learning manner. Typical works include CLIP \cite{radford2021learning}, etc. Some researchers have focused on modeling multi-modal user sequences in recommendation. M5 \cite{10.1145/3580305.3599863} applies a multi-modal embedding layer to extract both ID embeddings of show ID and content-graph embeddings initialized from a meta-path pretrained model. To better increase the generalization of ID embeddings, some research is proposed to express item representations as quantized vectors in discrete codes, including Product Quantization \cite{5432202}, VQ-VAE \cite{10.5555/3295222.3295378}, RQ-VAE \cite{zeghidour2021soundstream}, etc. VQ-Rec \cite{hou2023learning} proposes to encode text as discrete codes using product quantization techniques and use transformer to learn cross-domain data in recommendation. TIGER \cite{rajput2023recommender} learns the semantic ID from the content information and learns RQ-VAE \cite{zeghidour2021soundstream} representations for generative retrieval.

Much research focuses on making fast retrieval of relevant items from an embedding database. Common methods include approximate nearest neighbor (ANN) search using HNSW \cite{malkov2018efficient}, Product Quantization \cite{5432202}, etc. Product quantization is a technique to transform a d-dimensional vector to a low-dimensional N-bit integer vector of centroid ids in the codebook. It first splits a vector $ \bold{x} \in \mathbb{R}^{D} $ into $N_{bit}$ sub-vectors and applies quantization function $ q(x)$ to assign each sub-vector $\bold{x}_{i}$ to the nearest centroid $c_{i}$ from a codebook $\mathcal{C}$ as $ \bold{x}=[\bold{x}_{i}]_{1:N_{bit}} \rightarrow [q(\bold{x}_{i})]_{1:N_{bit}}=[c_{i}]_{1:N_{bit}} $. And the quantization function is $q(\bold{x}_{i})=\arg \min_{c_{i} \in \mathcal{C}} d(\bold{x}_{i}, e_{c_{i}}) \label{eq:exact_attention} $ and $ e_{c_{i}} $ denotes the centroid embedding of the $ c_{i} $-th centroid.


\section{Proposed Model}

\subsection{Problem Formulation}

We can split the sequence of users' behaviors into several heterogeneous sub-sequences, including the sequence of search queries $\mathcal{Q}=\{q_{1},q_{2},..,q_{|\mathcal{Q}|}\}$ of explicit intents and the sequence of browsing recommended items $\mathcal{B}=\{i_{1},i_{2},...,i_{|\mathcal{B}|}\}$. For search behaviors, users input a query $q \in \mathcal{Q}$ and interact (click or view) with a few items related to the query, resulting in the aligned sequence of query and item pairs $(q_{l}, i_{l})$. For the behavior of browsing recommended items, users browse a sequence of items without explicit search intent, and we pad an empty search query $q=\emptyset$ to each item to obtain the query-item pair as $(q_{l}=\emptyset, i_{l})$. Finally, we construct a unified sequence of aligned query-item pairs in chronological order with length $L$, denoted as $\{(q_{l}, i_{l})\}_{l=1:L}$. In some recommendation scenarios, such as short video recommendations of YouTube and TikTok, each item has multi-modal features such as text (title of video), image, and attributes (authors and categories). We further split the sequence of browsed items $\mathcal{B}$ into $M$ multi-modal sub-sequences, including a sequence of text features $\mathcal{T}=\{T_{1},T_{2},...,T_{L}\}$, a sequence of image features $\mathcal{I}=\{I_{1},I_{2},...,I_{L}\}$, and a sequence of attribute features $\mathcal{A}=\{A_{1},A_{2},...,A_{L}\}$. Finally, we let $[\mathcal{Q},\mathcal{T},\mathcal{I},\mathcal{A}] \in \mathbb{R}^{(M+1)\times L \times d}$ denote the input sequence of multi-modal query-item pairs to the SEMINAR model and $d$ denotes the dimension of aligned representations.

\subsection{Aligned Lifelong Sequence of Multi-Modal Query-Item Pairs}

The aligned sequence of multi modal query-item pairs pass the embedding layers. We let $[\bold{x}_{l}=(\bold{x}^{query}_{l},\bold{x}^{item}_{l})]_{l=1:L}$ denote the historical sequence of query and item pairs. $\bold{x}^{query}_{l} \in \mathbb{R}^{d},\bold{x}^{item}_{l}=(\bold{x}^{text}_{l},\bold{x}^{image}_{l},\bold{x}^{attributes}_{l}) \in \mathbb{R}^{M \times d}$. In CTR prediction, target attention (TA) is a structure which uses target item to retrieve the most relevant items from the sequence of historical behaviors. We extend TA from target item to target query-item pair to retrieve most relevant top $K$ pairs from historical sequence. We denote the target query-item pair as $\bold{x}_{t}=(\bold{x}^{query}_{t},\bold{x}^{text}_{t},\bold{x}^{image}_{t},\bold{x}^{attributes}_{t})$. 

\subsection{SEMINAR Model Architecture}

Our proposed model SEMINAR in Figure \ref{fig:seminar_model_architecture} introduces a new network Pretraining Search Unit (PSU) to pretrain using dataset of the lifelong sequence of multi-modal query-item pairs. Section \ref{sec:psu} introduces the PSU and corresponding pretraining tasks. Section \ref{sec:finetune_ctr_prediction} introduces how the recommendation model restores the pretrained query and item representations from PSU as initialization and applies a projection matrix to get the transformed representation of the sequence. Top-K relevant pairs are retrieved by the target pair and participate in the multi-head target attention (MHTA) calculation. Section \ref{sec:pq_fast_retrieval} introduces the multi-modal product quantization approximation.



\subsubsection{Pretraining Search Unit}
\label{sec:psu}

The input to PSU is the aligned sequence of query-item pairs as $[\bold{x}_{l}]_{l=1:L}$. $L$ denotes the length of the aligned sequence and $\bold{x}_{l}=(\bold{x}^{query}_{l},\bold{x}^{text}_{l},\bold{x}^{image}_{l},\bold{x}^{attributes}_{l})$ represents the $l$-th behavior in the sequence, which consists of the query embedding and multi-modal embedding. The query $q \in \mathcal{Q}$ passes through the query feature encoder $f(.)$, resulting in $Q=f_{}(\bold{x}^{query}_{1:L}) \in \mathbb{R}^{L\times d_{query}}$. Following the multi-modal alignment literature such as CLIP \cite{radford2021learning}, we use Transformer \cite{vaswani2023attention} to encode the text feature as $T=\text{Encoder}_{text}(\bold{x}^{text}_{1:L}) \in \mathbb{R}^{L\times d_{text}}$ and ViT \cite{DBLP:journals/corr/abs-2010-11929} to encode the image features as $I=\text{Encoder}_{image}(\bold{x}^{image}_{1:L}) \in \mathbb{R}^{L\times d_{image}}$. Additionally, we encode features of the attributes using the function $g(.)$. $A=g(\bold{x}^{attribute}_{1:L}) \in \mathbb{R}^{L\times d_{attribute}}$ is treated as one channel of the sub-sequence which participates in the multi-modal alignment of the item sequence. To project the representations of different channels to the same dimension $d$, we further multiply them by linear weight matrix $ \{W_{q},W_{t},W_{i},W_{a}\}$  and get the stacked input sequence of multi-modal query-item pairs as follows: $ \bold{x}=[QW_{q},TW_{t},IW_{i},AW_{a}] \in \mathbb{R}^{(M+1) \times L \times d} $.

\paragraph{\textbf{Next Pair Prediction and Multi-Head Target Attention}}

The intuition behind PSU is to design a pretraining network to learn from the lifelong behavior sequence, and the pretraining network should share the same structure of multi-head target attention with the cascading two-stage downstream model, such as ETA \cite{chen2022efficient} and TWIN \cite{10.1145/3580305.3599922}. The downstream model restores the pretrained query and item embeddings as initialization of parameters and fine-tunes the network. Different from the masked language model (MLM) in BERT \cite{devlin2019bert}, which uses tokens from the context window to predict the masked token, we use next-pair prediction as a pretraining task to predict the correct last query and item pair. We intentionally leave out the last query-item pair in the sequence $\bold{x}_{L}=[\bold{x}^{(m)}_{L}], m \in [Q,T,I,A]$, and treat it as the target query-item pair to retrieve from the previous $(L-1)$ sequence using multi-head target attention. To pad the sequence length from $L-1$ to $L$, we further add a special token $<EOS>$ to the end of the previous $L-1$ items in the sequence $\bold{x}_{1:L-1}$. The next query-item pair prediction task is formulated as classification tasks: $ y=p(\bold{x}^{1:M+1}_{L}|\bold{x}^{1:M+1}_{1:L-1};\bold{x}^{<EOS>})$ with the loss $\mathcal{L}^{pair}_{next}$. Positive label is assigned to the correct last pair, and negative labels are assigned to negatively sampled query-item pairs.

To better represent the historical behaviors and target query-item pair, we need to fuse the query and multi-modal item representations into a single vector as: 

\vspace{-1.0em}

$$\bold{x}=\lambda \bold{x}^{query} + (1-\lambda) \sum_{i} w_i{\bold{x}^{item}}^{(i)}=\sum_{m \in M+1} \gamma_{m}\bold{x}^{(m)}\label{eq:item_fusion}$$
\vspace{-1.0em}

$\lambda$ and $(1-\lambda)$ denote the weight to merge representations of query and item vectors respectively as $\lambda \in [0,1]$, and $w_{i}$ denotes the weight to merge multi-modal item representations. To simplify the notations, we use a single vector $[\gamma_{m}]^{1:M+1} \in \mathbb{R}^{M+1}$ to represent the weight of all $(M+1)$ channels and the sum of the weight equals to 1 as $\sum \gamma_{m}= 1$. The weight vector $\gamma_{m}$ can be learned dynamically as the softmax output of a gating network. The attention is calculated as the inner product of queries and keys of the merged multi-channel representations. The final attention score will be dominated by the modals with large norm values $|x^{(m)}|$ and large weight $\gamma_{m}$, and the information from other modals will be easily ignored. So we specifically decompose the attention score calculation into the norm value part $|x^{(m)}|$ and unit vector part $\hat{x}^{(m)}$. 

We let $ q_{t}$ denote the representation of target query-item pair as $q_{t}=\sum_{i}\gamma_{i}\bold{x}^{(i)}_{t}=\sum_{i}\gamma_{i}|\bold{x}^{(i)}_{t}|\hat{\bold{x}}^{(i)}_{t}$. Note that the $|\bold{x}^{(i)}_{t}|$ denotes the norm value of the i-th channel of target item and $\hat{\bold{x}}^{(i)}_{t}$ is a unit vector. Similarly, we can express the $l-th$ historical behavior $k_{l} \in K$ as $k_{l}=\sum_{j}\gamma_{j}\bold{x}^{(j)}_{l}=\sum_{i}\gamma_{j}|\bold{x}^{(j)}_{l}|\hat{\bold{x}}^{(j)}_{l}$. Note that the unit vectors of multi-modal sequence representations will participate in the multi-modal alignment task in the next section.

The $h$-th head in the multi-head attention is represented as $head^{PSU_{h}}=\text{Attention}_{h}(q_{t},K^{PSU},V^{PSU})$, and the attention score $a^{PSU}_{h}$ is calculated as inner product of d-dimensional vector query and keys 
multiplied by a scaling factor $ \frac{1}{\sqrt{d}} $.


$$ \alpha_{h}^{PSU} =\frac{(q_{t}W^{PSU_{Q}}_{h})(K^{PSU}W^{PSU_{K}}_{h})^{T}}{\sqrt{d}} $$

$$=[\sum_{i}\sum_{j}\gamma_{ij}(\bold{\hat{x}}^{PSU(i)}_{t}W^{PSU_{Q}}_{h})({\bold{\hat{x}}^{PSU(j)}_{l}}W^{PSU_{K}}_{h})^{T}]^{L}_{l=1} $$

$$ \gamma_{ij} = \gamma_{i}\gamma_{j} |\bold{x}^{PSU(i)}_{t}||\bold{x}^{PSU(j)}_{l}| $$

In this formulation, $ [\bold{x}^{ PSU(1:M+1)}_{l}]^{L}_{l=1} \in \mathbb{R}^{L\times (M+1) \times d} $ denotes the multi-modal  embedding of items in the sequence of PSU and $ \bold{x}^{PSU(1:M+1)} = [Q,T,I,A] $. And $K^{PSU}=[\sum_{i}\gamma_{i}\bold{x}^{PSU(i)}_{l}]^{L}_{l=1}\in\mathbb{R}^{L\times d}$ denotes the merged representations of input sequence. $W^{PSU_{Q}}_{h} \in \mathbb{R}^{d\times d}$ and $W^{PSU_{K}}_{h} \in \mathbb{R}^{d\times d}$ denote the projection weight matrix of query and keys in $h$-th head, and $\gamma_{ij}$ denotes the weight of cross-modal interaction of unit query vector and unit key vector in the sequence. $ \gamma_{ij} $ equals to the scalar product of $\gamma_{i}$, $\gamma_{j}$, the norm value of query vector $|\bold{x}^{PSU(i)}_{t}|$ and the norm value of key vector $|\bold{x}^{PSU(j)}_{l}|$.


\paragraph{\textbf{Multi-Modal Alignment and Query-Item Relevance}}

Multi-modal alignment is a crucial task, which learns the multi-modal representation in a same embedding space. Typical alignment models, such as CLIP \cite{radford2021learning}, maximize the cosine similarity of the correct $N$ (text-image) pairs and minimize the cosine similarity of the incorrect $N^2-N$ mismatch pairs. 
We simultaneously train multi-modal alignment tasks, including text-image, image-attributes, text-attributes with the cross entropy loss of N pairs. 


$$ \mathcal{L}_{align}=\sum_{i \in M}\sum_{j \in M \ne i}\mathcal{L}_{\text{CLIP}}(\bold{\hat{x}}^{(i)}_{1:L},\bold{\hat{x}}^{(j)}_{1:L}), (i,j) \in \{T,I,A\} $$


Sequence length $L$ is usually large and the alignment has complexity of $O(L^{2})$. To reduce the complexity, we further split the sequence into $N_{ch}$ chunks. Each chunk is a sub-sequence with length $L_{sub}=\frac{L}{N_{ch}}$. The alignment loss is the sum of multiple losses within chunks as $ \mathcal{L}_{\text{CLIP}}(\bold{\hat{x}}^{(i)}_{1:L},\bold{\hat{x}}^{(j)}_{1:L})=\sum_{k \in N_{ch}} \mathcal{L}_{CLIP}(\bold{\hat{x}}^{(i)}_{L_{k}:L_{k+1}},\bold{\hat{x}}^{(j)}_{L_{k}:L_{k+1}}) $ with complexity reduced to $O(L^{2}/N_{ch})$.

Additionally,  query item relevance prediction is a typical search task, usually modelled as binary classification to predict the correct query-item pair from irrelevant query-item pairs. Each pair of query and item is represented as $[\bold{x}^{query}_{l};\bold{x}^{item}_{l}=\sum_{m \in M}\gamma_{m}\bold{x}^{item^{(m)}}_{l}]$. Loss for query-item relevance binary classification task is $\mathcal{L}_{query-item}=\sum \mathcal{L}_{ce}(y^{qi}_{l};\bold{x}^{query}_{l},\sum_{m \in M}\gamma_{m}\bold{x}^{item^{(m)}}_{l})$. $ y^{qi}_{l} $ denotes relevance label of the l-th pair in the sequence. Positive label is assigned to the correct query-item pair and negative label is assigned to randomly sampled irrelevant query-item pair.

\vspace{-0.9 em}

\paragraph{\textbf{Loss of Pretraining Search Unit}}

The objective of Pretraining Search Unit (PSU) consists of three parts, the next query-item pair prediction loss $\mathcal{L}^{pair}_{next}$, multi-modal alignment loss $\mathcal{L}_{align}$ and the query-item relevance prediction loss $\mathcal{L}_{query-item}$. $\mathcal{L}_{PSU}=\mathcal{L}^{pair}_{next}+\mathcal{L}_{align}+\mathcal{L}_{query-item}$. 

\subsubsection{Fine-tuning the projection weight}
\label{sec:finetune_ctr_prediction}

Existing lifelong sequence modeling methods follow a cascading two-stage paradigm. In the first stage, target item or query is used as trigger to retrieve the most relevant top-K items from the users' long behaviors sequence and reduce the sequence length from $L$ to $K$, such as the General Search Unit (GSU) in SIM \cite{10.1145/3340531.3412744}, TWIN \cite{10.1145/3580305.3599922}, and Relevance Search Unit (RSU) in QIN \cite{10.1145/3583780.3615022}. In the second stage, a multi head target attention (MHTA) unit in Exact Search Unit (ESU) is applied to encode the selected $K$ relevant items as the representation of users' behavior sequence. However, existing cascading two-stage paradigm suffers from the insufficient learning problem of ID embedding in the lifelong sequence ($[\bold{x}^{query},\bold{x}^{text},\bold{x}^{image},^{attributes}]$). The downstream model, e.g. CTR prediction, lacks of enough training data to learn the embedding in the sequence well. Especially when some low-frequency items in the sequence exist a long time ago (more than one year) and don't exist in the training data, which are collected from most recently users' logs. 

To help alleviate the insufficient learning problem of ID embedding in the lifelong sequence, the general search unit (GSU) in our proposed SEMINAR model shares the same multi-head target attention structure $head^{GSU_{h}}=\text{Attention}_{h}(q_{t}, K^{GSU}, V^{GSU})$ with the structure in PSU as $head^{PSU_{h}}=\text{Attention}(q_{t}, K^{PSU}, V^{PSU})$, restores the pretrained embedding from PSU and applies specific projection weight matrix $G^{(j)} \in \mathbb{R}^{d \times d}$ to the pretrained embedding. After the first stage retrieval, the sequence length is reduced from $L$ to $K$, the second stage ESU also shares the same multi-head target attention structure $head^{ESU_{h}}=\text{Attention}(q_{t}, K^{ESU}, V^{ESU})$ with GSU and PSU, and has specific projection weight matrix $W^{Q}_{h},W^{K}_{h},W^{V}_{h}$ of each head. 

GSU restores the pretrained query item multi-modal embedding $[E^{PSU(Q)},E^{PSU(T)},E^{PSU(I)},E^{PSU(A)}] $ from PSU, and applies projection matrix $G^{(j)} \in \mathbb{R}^{d \times d}$ to get the projected embedding in GSU as $\bold{x}^{GSU(j)}$. $ E^{PSU(*)} $ denotes the pretrained multi-modal embedding. And the attention score $ \alpha_{h}^{GSU} $ in the $h$-th head of GSU's multi-head target attention is calculated as: 
$$ \alpha_{h}^{GSU}=\frac{(q_{t}W^{GSU_{Q}}_{h})(K^{GSU}W^{GSU_{K}}_{h})^{T}}{\sqrt{d}} $$
$$ \bold{x}^{GSU(j)}=\bold{x}^{PSU(j)}G_{j}, \forall j \in M+1 $$



Comparing the GSU attention $\alpha^{GSU}_{h}$ with the pretrained PSU attention $\alpha^{PSU}_{h}$, we can see that the structures of multi-head target attention are exactly the same. The projection weights $W^{GSU_{Q}}_{h}$ and $W^{GSU_{K}}_{h}$ of queries and keys for each head in multi head attention are different from $W^{PSU_{Q}}_{h}$ and $W^{PSU_{K}}_{h}$. And the embedding projection weight matrix $G_{j}$ is unique to GSU.


In the second stage, the top-K relevant query-item pairs are selected from GSU and fed to Exact Search Unit (ESU) as $head^{ESU_{h}} = \text{Attention}_{h}(q_{t},K^{ESU},V^{ESU})$.

In ESU, $K^{ESU}  =\text{TopK}(K^{GSU}) \in \mathbb{R}^{(M+1) \times K \times D} $ represents the sequence of retrieved top-K representations from $K^{GSU} \in \mathbb{R}^{(M+1) \times L \times D}$. The attention score in ESU is denoted as $\alpha^{ESU}_{h}$ and the ID embedding in ESU is denoted as $\bold{x}^{ESU(j)}$. 

$$\alpha^{ESU}_{h}=\frac{(q_t W^{ESU_{Q}}_{h})(K^{ESU} W^{ESU_{K}}_{h})^{T}}{\sqrt{d}}$$

$$\bold{x}^{ESU(j)}=\bold{x}^{GSU(j)}=\bold{x}^{PSU(j)}G^{(j)}, \forall j \in M+1$$.

Finally, users' lifelong sequence representation $\bold{x}_{\text{lifelong\_seq}}$ is calculated as: $\bold{x}_{\text{lifelong\_seq}} = \text{Concat}(\text{head}^{ESU_{1}},...,\text{head}^{ESU_{H}}) W^{ESU} $. And $\bold{x}_{\text{lifelong\_seq}}$ is concatenated with other user, item, user-item interaction (u2i) and context features and participate in CTR prediction. $\hat{y}_{i}=f_{\theta_{i}}(\bold{x}_{\text{lifelong\_seq}},\bold{x}_{u},\bold{x}_{i},\bold{x}_{\text{u2i}},\bold{x}_{\text{context}})$ denotes the predicted value and $y_{i}$ denote the actual label value. And the final loss of CTR prediction is $\mathcal{L}_{ctr}=\sum_{i} \mathcal{L}_{ce}(y_{i}, \hat{y_{i}})$.


\subsubsection{Approximate Retrieval of Multi-Modal Query-Item Pair}
\label{sec:pq_fast_retrieval}
The exact calculation of the attention score between the target query-item pair $q_{t}$ and the $l$-th query-item behavior $k_{l}$ is the inner product of the weighted sum of multiple vectors as:
$${q_{t}}^{T}k_{l}=(\sum_{i \in M+1}\gamma_{i}{x}^{(i)}_{t})^{T}(\sum_{j \in M+1}\gamma_{j}{x}^{(j)}_{l}), \forall l \in \{1,2,...L\}$$

 The exact calculation has the time complexity of $ O(L\times M \times d) $. $L$ denotes the sequence length, $M$ denotes the number of weighted sum operations of multi-modal embedding vectors of dimension $d$. The calculation becomes time-consuming when $L$ is very large ($10^4$) in the lifelong sequence of multi-modal query-item pairs setting. 


One straightforward method of fast retrieval $K$ nearest vectors given an input query vector $q$ is to build an embedding index, such as HNSW \cite{malkov2018efficient}, and conduct ANN (Approximate Nearest Neighbors) search. However, there are difficulties in building an embedding index to retrieve the target query-item pair from the sequence of multi-modal query-item pairs. To search the vectors of behaviors given the input target query-item pair $q_{t}$ as in the exact attention calculation, we build a vector index which assigns a primary key to represent each vector, such as Item ID, Query ID, etc. However, in our aligned sequence of query-item pairs, each merged query-item representation have the joint key of (query\_id, item\_id), and the required amount of storage increases from the item set size $|\mathcal{B}|$ to the cartesian product of the query set size $|\mathcal{Q}|$ and the item set size $|\mathcal{B}|$ as $|\mathcal{Q}||\mathcal{B}|$, which is almost infeasible to store the merged query-item pair in a single index directly.

An alternative cascading cross-modal strategy is considered to retrieve top-K relevant query-item pairs. Firstly, we build two separate vector indexes of the query set with size $|\mathcal{Q}|$ and the item set with size $|\mathcal{B}|$. During the online retrieval of target query-item pairs, we conduct vector retrieval four times, including query-to-item, query-to-query, item-to-query, and item-to-item. Each retrieval keeps the top-K items with the maximum inner product. The filter in the first-stage cross-modal retrieval is $L \rightarrow 4K$. Given the potential $4K$ items, we conduct an exact attention calculation on these items to obtain the final top-K items, and the filter is $4K \rightarrow K$.
The problem with the cascading cross-modal retrieval strategy is that it may achieve a suboptimal solution compared to exact full attention calculation. This is because the final inner product is a weighted average of all modalities. Additionally, top-K relevant items from one modality (e.g., query-to-query relevance) may have very low relevance in other modalities, such as query-to-item (text) or query-to-item (image), thus the overall inner product score is not optimal. $Recall@K$ can evaluate the performance of the greedy strategy compared to exact calculation.

To help increase the recall performance while considering the retrieval speed, the key is to reduce the cardinality of the query set $\mathcal{Q}$ and the item set $\mathcal{B}$. We argue that product quantization is a good approximation strategy, which splits vectors into $N_{bit}$ sub-vectors, assigns each sub-vector to the nearest centroid, and reduces the cardinality. In our formulation, we first use a set of separate $N_{bit}$ quantization function $[q^{(m)}_{1},q^{(m)}_{2},…,q^{(m)}_{N_{bit}}]$ to encode embedding of vectors from the $m$-th modal channel $\bold{x}^{(m)}$ as integer vectors of ${N_{bit}}$-dimension, $ q({\bold{x}}^{(m)})=[c^{(m)}_{1},c^{(m)}_{2},...,c^{(m)}_{N_{bit}}] \in \mathbb{R}^{N_{bit}} $. Each representation of multi-modal query-item pair is expressed as:
\vspace{-0.5em}
$$ [{\bold{x}}^{(1)},...,{\bold{x}}^{(M)}] \rightarrow [q({\bold{x}}^{(1)}),...,q({\bold{x}}^{(M)})] \in \mathbb{R}^{M \times N_{bit}} $$.
\vspace{-1.0em}

We pre-compute the inner product between different pairs of centroids and store the values in memory. The space complexity of the storage is $ O(M^{2}|\mathcal{C}|^{2}N_{bit}) $, where $M$ denotes the size of multi-modals, $|\mathcal{C}|$ denotes the number of centroids, and $N_{bit}$ denotes the number of subvectors split in the codebook of modal $m$. During online serving, the inner product 
 of $ {q_{t}}^{T}k_{l} $ is equivalent to $O(M^{2}N_{bit})$ distance lookup operations, and the final score is calculated as the weighted sum of these distances. Here, $ c^{(i)}_{b} $ and $ c^{(j)}_{b} $ denotes the centroids IDs of the $b$-th subvector of $ \bold{x}^{(i)}_{t} $ and $ \bold{x}^{(j)}_{l} $ respectively.

\vspace{-1.0em}
$$ {q_{t}}^{T}k_{l}= \sum_{i}\sum_{j}\gamma_{i}\gamma_{j}\bold{x}^{(i)}_{t}\bold{x}^{(j)}_{l}\approx\sum_{i}\sum_{j}\gamma_{i}\gamma_{j}\sum_{b \in N_{bit}}\text{dist}(c^{(i)}_{b},c^{(j)}_{b}) $$
\vspace{-1.0em}

Our proposed multi-modal product quantization strategy works quite well in real-world settings. We also compare the time complexity of different strategies, such as cascading ANN (HNSW), Locality-sensitive hashing (LSH) and our proposed Multi-Modal Product Quantization approximation. Our proposed multi-modal PQ method has the time complexity of $ O(L \times M^{2} \times N_{bit}) $. In each attention calculation, there are $ M^{2}N_{bit} $ distance look-up operations of $O(1)$, and the final score is calculated as the sum of these distances, which is far less than the exact calculation of the inner product of multiple vectors $ O(L\times M \times d) $. As for the two stage cascading ANN (HNSW) method of retrieving query-item pairs with two filters, the first stage retrieve the $M^{2}K$ cross-modal candidates from $L$ sequence as $L \rightarrow M^{2}K$, and the second stage retrieve the final top $K$ items from first stage as $M^2K \rightarrow K$. Total time complexity of cascading ANN method is $ O(M^{2}\log(L)d+M^{2}Kd) $, which is faster than our PQ strategy but may achieve sub-optimal recall performance in multiple experiments as reported in Figure \ref{fig:seminar_approximate_retrieval}.

\section{Experiment}
\label{sec:exp}
\subsection{Experimental Settings}

\textbf{Dataset}
We evaluate our proposed SEMINAR model on three datasets: two public datasets including Amazon review dataset (Movies and TV subset) \footnote{\url{https://cseweb.ucsd.edu/~jmcauley/datasets/amazon\_v2/}} and the KuaiSAR \footnote{\url{https://zenodo.org/records/8181109}} search and recommendation dataset, one industrial dataset Alipay short video dataset. The average length of users' sequence has the magnitude of $L=2000,1000,100$ for the Alipay, KuaiSAR and Amazon datasets. The detailed statistics can be found in Table \ref{tab:statistics}.

\begin{itemize}
\item{\textbf{Amazon Reviews}}
We select Movies and TV subset of the public Amazon reviews dataset for experiment. The meta-information of items is also provided in the dataset. We use the image thumbnails as the inputs to the sequence of image modal. To get the aligned query sequence, we generate a query relevant to each item from its description in the meta information as in \cite{10.1145/3077136.3080813} and \cite{10.1145/3583780.3615022}. 

\item{\textbf{KuaiSAR} \cite{kuaisar_dataset}}
KuaiSAR is a real-world public large scale dataset containing both search and recommendation behaviors collected from Kuaishou\footnote{https://www.kuaishou.com/en}, a leading short-video app. We construct a unified sequence of query and item pairs to compare different lifelong behaviors sequences models. 

\item{\textbf{Alipay Short Video}}
The Alipay short video dataset is a real-world industrial dataset collected from exposures and clicks logs of short-video recommendation and search ranking scenario of Alipay app. We convert the title of the short video as the input to the text modal, and the image thumbnails to the image modal. Users' search queries are collected and aligned to corresponding viewed items.

\end{itemize}



We process the datasets of Amazon and KuaiSAR as in literature \cite{S_3_rec_ssl} and repo \footnote{\url{https://github.com/RUCAIBox/CIKM2020-S3Rec}}. User with $N$ actions will generate N-1 samples. We use the first $i-1$ actions to predict whether the user will interact with the $i$-th item ($0<i<=N$). Additionally, we apply the leave-one-out strategy, using the $(N-1)$-th action as the validation set, the $N$-th action as positive in test set and randomly sampled negatives in the test set. The remaining samples are used as training and pretraining set. In the industrial Alipay short video dataset, exposed clicks are treated as positive samples and exposed non-clicks are considered as negative samples. The training and validation sets are randomly split using data from past [0,T-1] days (T=60), and the test set come from the $T$-th day.

\begin{table}
  \caption{Statistics of the Amazon Movies and TV, the Alipay Short Video and KuaiSAR datasets. K denotes thousand.}
  \label{tab:statistics}
  \begin{tabular}{|c|c|c|c|c|}
    \hline
    Dataset & User & Item & Query & U-I \\
    \hline
    Amazon Movies \& TV & 297 K & 181 K & - & 3,293 K \\
    \hline
    Alipay Short Video & 35,065 K & 1,132 K & 51 K & 62,948 K \\
    \hline
    KuaiSAR & 25,877 & 6,890,707 & 453,667 & 19,664,885 \\    
    \hline
\end{tabular}
\end{table}

\begin{table*}
  \caption{Results of lifelong behavior sequence modeling of KuaiSAR dataset, Amazon Review dataset and Alipay short video recommendation dataset.* indicates best performing model. }
  \label{tab:experiment_result_amazon}
  \centering
  \begin{tabular}{|c|c|c|c|c|c|c|c|}
    \hline
     & \multicolumn{3}{|c|}{KuaiSAR} & \multicolumn{3}{|c|}{Amazon Movies and TV} & Alipay Short Video\\
    \hline
    Method& NDCG@5 & NDCG@10 & NDCG@50 & NDCG@5 & NDCG@10 & NDCG@50 & AUC \\
    \hline
    SIM & 0.2523 & 0.2661 & 0.3293 & 0.3573 & 0.3959 & 0.4577 & 0.7382 \\
    \hline
    QIN	& 0.2535 & 0.2672 & 0.3312 & 0.3650 & 0.4038 & 0.4630 & 0.7239 \\
    \hline
    ETA & 0.2642 & 0.2756 & 0.3313 & 0.3626 & 0.4008 & 0.4607 & 0.7262 \\
    \hline
    TWIN & 0.2558 & 0.2709 & 0.3294 & 0.3627 & 0.4017 & 0.4605 & 0.7376 \\
    \hline
    SEMINAR & *0.2816 & *0.2969 & *0.3457 & *0.3661 & *0.4041 & *0.4636 & *0.7503\\
    \hline
    Absolute Impr. & +0.0292 & +0.0308 & +0.0164 & +0.0088 & +0.0082 & +0.0059 & +0.0264\\
    \hline
  \end{tabular}
\end{table*}

\begin{table}
  \caption{Ablation Studies of Different PSU pretraining tasks on KuaiSAR Dataset. N@K denotes $\text{NDCG@K}$.}
  \label{tab:experiment_result_psu_abalation}
  \begin{tabular}{|c|c|c|c|}
    \hline
    Method & N@5 & N@10 & N@50 \\
    \hline
    SEMINAR	& 0.2816 & 0.2969 & 0.3457 \\
    \hline
    w/o pretraining & 0.2564 & 0.2738 & 0.3310 \\
    \hline
    w. align, w/o next-predict,q-i relev. & 0.2702 & 0.2832 & 0.3420 \\
    \hline
    w. next-predict, w/o align,q-i relev. & 0.2675 & 0.2813 & 0.3408 \\
    \hline
    w. q-i relev., w/o align, next-predict & 0.2633 & 0.2754 & 0.3357 \\
    \hline
  \end{tabular}
\end{table}

To evaluate the recall performance of different approximation fast retrieval methods, we conduct experiments on two datasets: the multi-modal embedding of the Alipay short video dataset with sequence length $L=2,000$ and a synthetic dataset. The purpose of the synthetic dataset is to test the performance of different retrieval methods on extremely long sequence (e.g. $L=10,000$), which is not available in public datasets. The synthetic dataset consists of query, text, image, and attribute vectors generated by i.i.d. normal distribution $N(\mu,\sigma^{2})$ with different values of mean $\mu$ and variance $\sigma^{2}$, to imitate various norm values of multi-modal vectors of query-item pairs.

\begin{table}
  \caption{$\text{Recall@K}$ Evaluation of Approximate Retrieval Methods on Alipay Short Video Recommendation Dataset}
  \label{tab:alipay_short_video_recall}
  \begin{tabular}{|c|c|c|c|c|}
        \hline
        Method & R@32  & R@64 & R@128 & R@256 \\
        \hline
        ANN (HNSW) & 0.7881 & 0.8603 & 0.9288 & 0.9409 \\
        \hline
        LSH & 0.7528 & 0.8175 & 0.8721 & 0.9257 \\
        \hline
        RQ-VAE  & 0.8225  & 0.8422  & 0.8633  & 0.8995 \\
        \hline
        Multi-Modal PQ & 0.9638 & 0.9769 & 0.9797 & 0.9874 \\
        \hline
\end{tabular}
\end{table}

\textbf{Comparison Methods}

We compared several strong lifelong sequence modeling baselines with our proposed SEMINAR model:

\begin{itemize}
\item{\textbf{SIM} \cite{10.1145/3340531.3412744}}
SIM adopts cascading search unit GSU and ESU to extract the relevant behaviors of the candidate item and applies multi-head target attention to model users’ interest.

\item{\textbf{ETA} \cite{chen2022efficient}}
Efficient Target Attention encodes query and keys as binary hash vectors using a multi-round random projection matrix. The retrieval is calculated as the Hamming distance between the target item and the items in the sequence.

\item{\textbf{TWIN} \cite{10.1145/3580305.3599922}}
Two-Stage Interest Network adopts the same relevance metric between the target behavior and historical behaviors as the target attention in two cascading stages GSU and ESU.

\item{\textbf{QIN} \cite{10.1145/3583780.3615022}}
QIN network uses the query as first trigger to retrieve top $K1$ behaviors, and target item as the second trigger to retrieve top $K2$ relevant items afterwards.
\end{itemize}

The input features to all baseline models are the same, including the query and multi-modal item features in all datasets.

For the online approximate retrieval performance, we compared our proposed multi-modal product quantization \cite{5432202} strategy with some widely adopted vector retrieval methods in the query-item multi-modal pairs retrieval setting, including:

\begin{itemize}
\item{\textbf{HNSW} \cite{malkov2018efficient}}
Navigable Small World Graphs
\item{\textbf{LSH}}
Locality Sensitive Hashing
\item{\textbf{RQ-VAE} \cite{zeghidour2021soundstream}}
Residual Vector Quantization VAE (RQ-VAE) follows an encoder-decoder structure and uses the multi-stage vector quantizer to regress original inputs, with multi-scale spectral reconstruction loss as constraints.

\end{itemize}

\textbf{Evaluation Metrics}
We use $\text{NDCG@K}$ to evaluate the recommendation performance on Amazon review and KuaiSAR dataset, and $\text{AUC}$ (Area Under the Curve) to evaluate the CTR prediction performance of exposures and clicks in Alipay short video dataset. Secondly, to evaluate the performance of multi-modal query-item retrieval, we calculate the exact attention on all the items in the sequence using the target query-item pair as a trigger and regard the real top-K relevant items as ground truth. Different fast retrieval strategies are evaluated by $\text{Recall@K}$ at different $K$ levels, which measures how many top-K relevant ground truth items are recalled by the approximation strategy.

\textbf{Implementation Details}
\label{sec:implementation}
We implement the baseline methods and our proposed SEMINAR model using PyTorch. Secondly, the baselines of different approximate retrieval methods, ANN(HNSW), and our Multi-Modal Product Quantization are implemented using the Python library faiss \cite{douze2024faiss} \footnote{\url{https://github.com/facebookresearch/faiss}}, and the RQ-VAE \cite{zeghidour2021soundstream} is implemented using the Python library vector\_quantize\_pytorch \footnote{\url{https://github.com/lucidrains/vector-quantize-pytorch}}.
The code of SEMINAR is available at the repo: \url{https://github.com/paper-submission-coder/SEMINAR} and the public datasets Amazon and KuaiSAR can be downloaded following the instructions in the README file.

For the hyperparameter settings of recommendation models, we set the sequence length $L$ to 2000,1000 and 100 for Alipay Short Video, KuaiSAR, and Amazon Review datasets and retrieve $K=200,200,50$ most relevant items, based on users' average interaction length in different datasets. The embedding of multi-modal text and image channels are outputs from pretrained ViT-B/32 \footnote{\url{https://github.com/openai/CLIP}} model of CLIP with original dimension 512, then linearly projected to dimension 64. And the weight of query representation $\lambda$ in section \ref{eq:item_fusion} is set to 0.5 to fuse query embedding and multi-modal item embedding. We also compare different $\lambda$ values ($\lambda=0.1,0.3,0.5,0.7,0.9$) in the following section of ablation study. For the multi-head target attention, we set number of heads as 4. The batch size is set to 256 and we are using Adam optimizer with learning rate set to 0.001. The number of pretraining epochs is set to $5, 1, 1$ on KuaiSAR, Amazon and Alipay datasets respectively, and the number of training epochs are the same for all models in comparison. The checkpoint is exported by best NDCG metrics on evaluation dataset. For the implementation of our multi-modal production quantization, number of modals $M$ is set to 4. And the original 64-dimension dense embedding vectors are expressed as $N_{bit}=8$ bits vectors of integer codes. Each bit of the integer vectors represents the codebook assignment of centroids $c^{(m)}_{i} \in \{1,2,…,|\mathcal{C}_{m}|\}$. The cardinality of each dimension $|\mathcal{C}_{m}|$ is set to 512. To generate the synthetic dataset with extremely long sequence $L=10,000$, We generate multi-modal embedding of sequence with different norm values across modals as normally distributed variables $N(\mu,\sigma^2)$. We set $\mu=0.25,0.5,1.0,2.0$ and $\sigma=1.0$ to query, attribute, text and image modals respectively. To investigate the influence of different fusion weight $\gamma_{m}$ of multi-modal embedding, we conduct different experiments of equal weights $\gamma_{m}=[0.25,0.25,0.25,0.25]$ and different weights $\gamma_{m}=[0.1,0.2,0.3,0.4]$.

\subsection{Experimental Results}

\begin{figure*}
  \includegraphics[height=4.42in, width=7.2in]{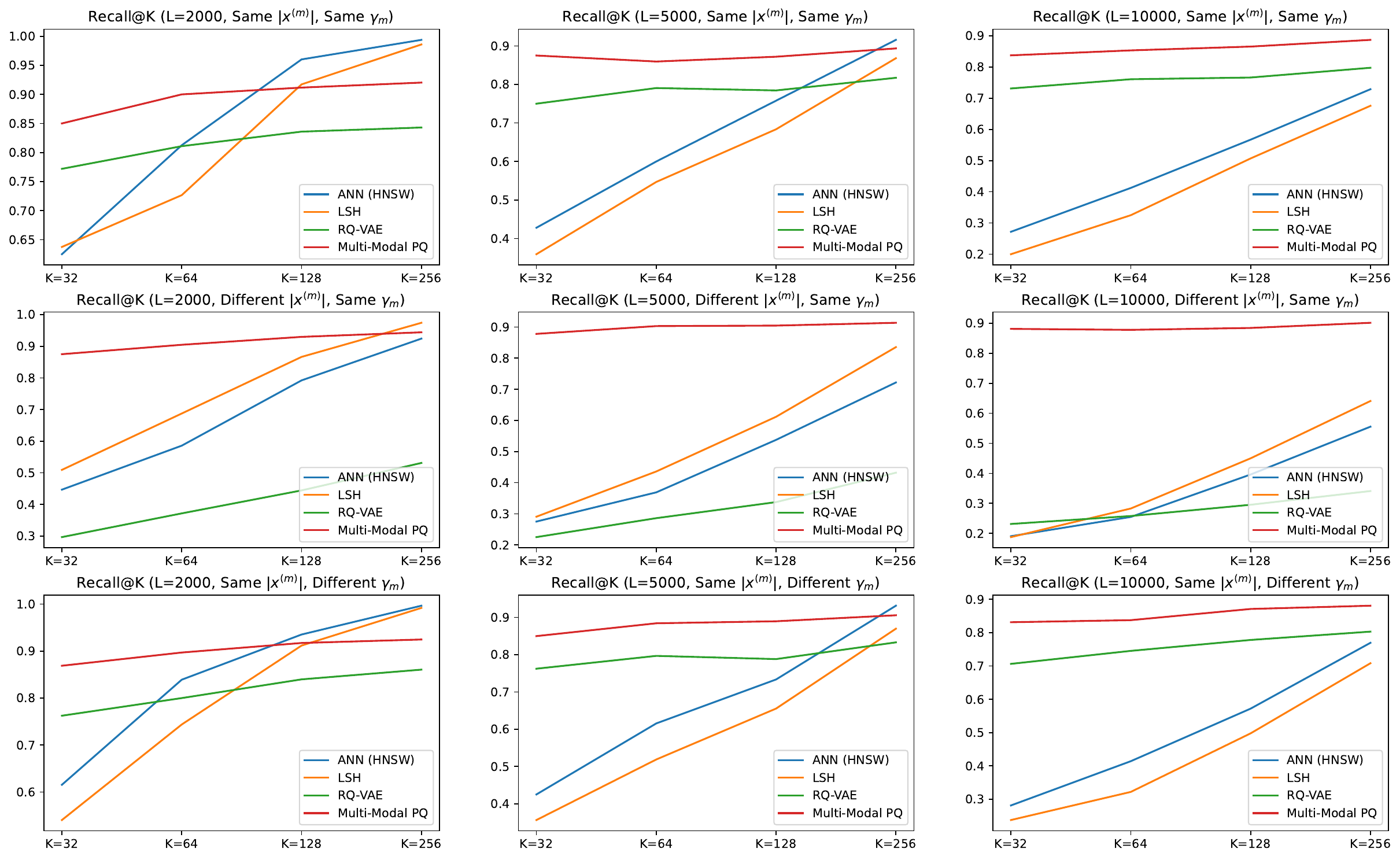}
  \caption{$\text{Recall@K}$ Evaluation of Different Approximate Fast Retrieval Methods on Synthetic Dataset of the Multi-Modal Lifelong Sequence. Plots in the first row denote the group of same norm $|x^{(m)}|$ same weight $|\gamma_{m}|$, plots in the second row denote the group of different norm $|x^{(m)}|$ and same weight $|\gamma_{m}|$, and plots in the third row denote the group of the same norm $|x^{(m)}|$ and different weight $|\gamma_{m}|$.}
  \label{fig:seminar_approximate_retrieval}
\end{figure*}

\subsubsection{Lifelong User Behavior Modeling}

We report the performance on different datasets from multiple domains, including $\text{NDCG@K}$ on KuaiSAR dataset, the Movie and TV subset of the Amazon review dataset and $\text{AUC}$ performance on the Alipay short video recommendation dataset in Table \ref{tab:experiment_result_amazon}. The asterisk (*) denotes the best performance achieved in each task. We can see that SEMINAR achieved the best performance on the KuaiSAR dataset with improvement of +0.0292, +0.0308, +0.0164 in $\text{NDCG@K}=5,10,50$ and improvement of +0.0088, +0.0082, +0.0059 in $\text{NDCG@K}=5,10,50$ on Amazon dataset compared to SIM. Additionally, SEMINAR also achieved the best $\text{AUC}$ performance on the Alipay short video recommendation dataset with improvement of +0.0264 compared to multiple strong SOTA baselines. 

\subsubsection{Multi-Modal Query-Item Pairs Approximate Retrieval}
To compare multi-modal query-item approximate retrieval methods, we report the $\text{Recall@K}$ performance of the industrial Alipay Short Video dataset in Table \ref{tab:alipay_short_video_recall} and the performance of the synthetic dataset in Figure \ref{fig:seminar_approximate_retrieval}. 
From the result of Alipay Short Video dataset, we observe that our proposed Multi-Modal Product Quantization strategy achieves the highest $\text{Recall@K}$ compared to other approximation methods under different values of $K=[32,64,128,256]$ and $L=2000$. Secondly, we observe that in the synthetic dataset, experimental groups are designed with different sequence length $L=2000,5000,10000$, different settings of norm values $|\bold{x}^{(m)}|$ across modalities, and different settings of weight $\gamma_{m}$ across modalities as in section \ref{sec:implementation} of detailed implementation. Our proposed method of multi-modal product quantization strategy consistently achieves the best $\text{Recall@K}$ at different $K$ levels ($K=[32,64,128,256]$), with only a few exceptions of falling behind the cascading ANN (HNSW) method for large $K$ values under $L=2000,5000$. For the first method, cascading ANN (HNSW), we observe that the greedy strategy of cascading ANN achieves poor results at small values of $K$ (e.g., L=10000, Recall@32=0.2719), and the performance increases dramatically as $K$ increases to 256 (L=10000, Recall@256=0.7289). This aligns with our expectation that in the setting of weighted sum of multiple vectors, as $K$ increases, the real top-K relevant pairs to the target pair have higher probability of being recalled by the greedy strategy of $M^2$ cascading cross-modal ANN retrieval.  

To analyze the effect of different variables for approximate retrieval, e.g. merging weights $\gamma_{m}$ of modalities, different norm values $|\bold{x}^{(m)}|$, we plot the line chart of $\text{Recall@K}$ as $K$ increases in Figure \ref{fig:seminar_approximate_retrieval}. From the chart, we can observe that different norm values of multi-modal vectors influence the overall $\text{Recall@K}$ dramatically. Under the same sequence length $L=10000$ and $K=256$, the $\text{Recall@K}$ of the group with different norm values is on average -0.14 below the group with the same norm values. The varied norm values of multi-modal vectors make it more challenging to achieve high $\text{Recall@K}$ compared to the equal norm values counterpart.

\begin{figure}
  \includegraphics[height=1.6in, width=3.2in]{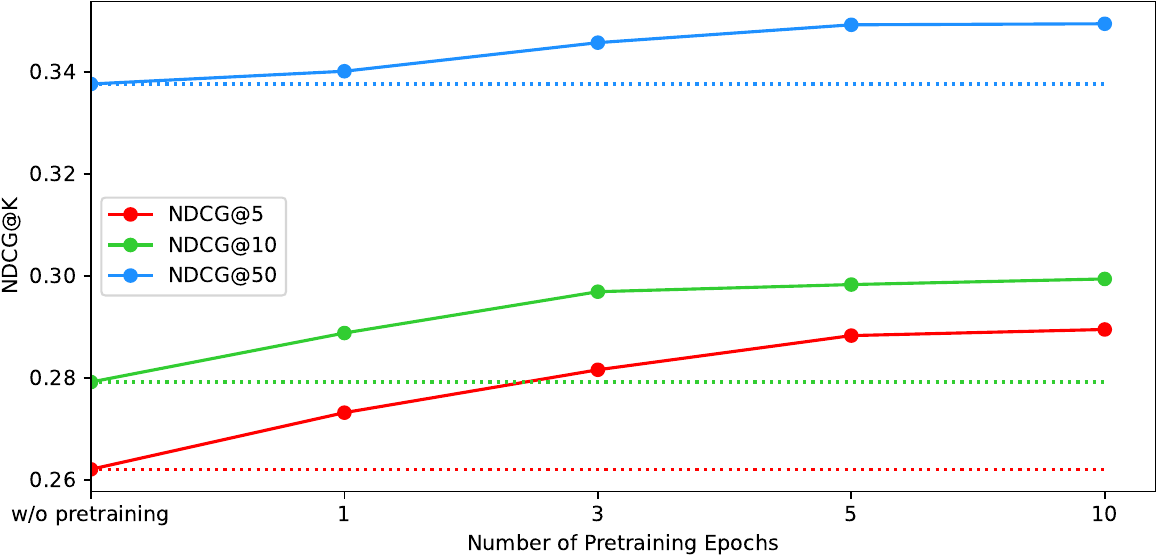}
  \caption{Influence of Number of Pretraining Epochs on $\text{NDCG@K}$ performance of KuaiSAR Dataset}
  \label{fig:ablation_pretrain_epochs}
\end{figure}

\begin{figure}
  \includegraphics[height=1.6in, width=3.2in]{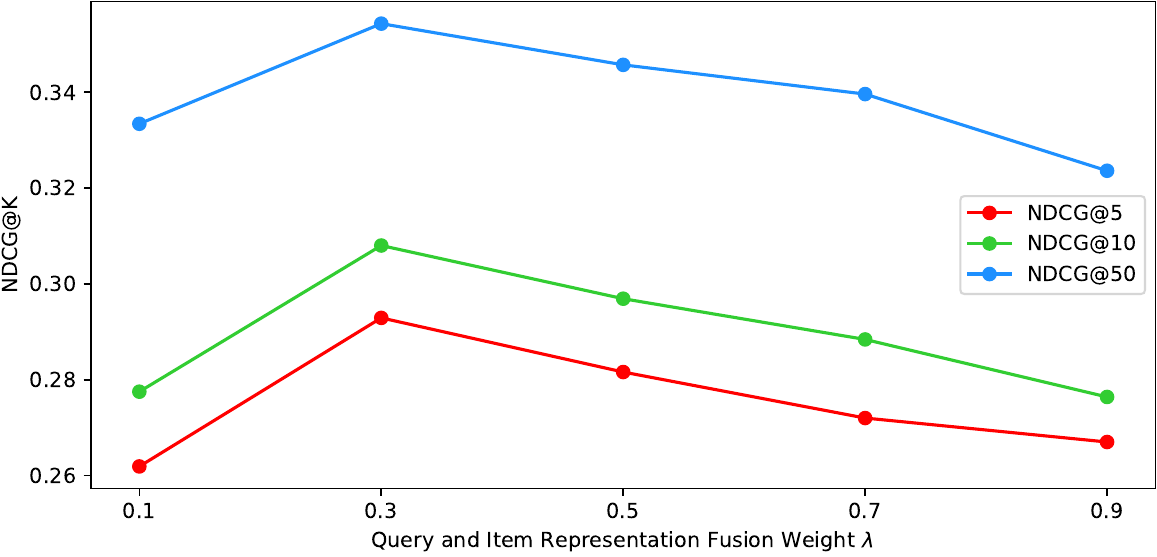}
  \caption{Influence of Query and Item Representation Fusion Weight $\lambda$ on \text{NDCG@K} performance of KuaiSAR Dataset}
  \label{fig:ablation_fusion_weight_lambda}
\end{figure}

\subsection{Discussion and Ablation Study}
\subsubsection{Influence of Pretraining Epochs Number and Ablation of Pretraining Tasks}
To investigate the influence of different pretraining epochs and without pretraining of the SEMINAR model, we reported the $\text{NDCG@K}$ performance on KuaiSAR dataset with sequence length 1000 in Figure \ref{fig:ablation_pretrain_epochs}. We can see that the SEMINAR model achieve largest improvement compared to the group of without pretraining in the first 5 pretraining epochs, and additional pretraining epochs up to 10 contribute only marginally to the performance. 

As for the ablation study of different pretraining tasks of SEMINAR, we trained different models on KuaiSAR dataset, including without pretraining, with only one pretraining task and without the other two tasks (e.g., w. alignment and w/o next pair prediction, query-item relevance). The results of the ablation study are reported in Table \ref{tab:experiment_result_psu_abalation}. Compared to the group of SEMINAR without pretraining, multi-modal alignment task contributes largest to the performance improvement, followed by next pair prediction and query-item relevance. 

\subsubsection{Query-Item Representation Fusion Weight}
To investigate the influence of different weight $\lambda$ to fusion query and item representation, we conduct different experiments $\lambda=[0.1,0.3,0.5,0.7,0.9]$ of SEMINAR model on KuaiSAR dataset. The results are reported in Figure \ref{fig:ablation_fusion_weight_lambda}. The best performance is achieved at $\lambda=0.3$. We speculate that optimal value of fusion weight $\lambda$ depends on the distribution of search and recommendation behaviors in the unified sequence of query-item pair. For example, in KuaiSAR dataset search actions consist of 25.7\% of overall users’ actions, and recommendation actions consist of 74.3\% of total actions as in \cite{kuaisar_dataset} . The optimal value of $\lambda$ may vary across different domains and datasets, which need further investigation in future research. 

\section{Conclusion}

In this paper, we proposed SEMINAR to model users' lifelong behavior sequence of query and item pairs. We introduced the Pretraining Search Unit to help alleviate the issues of insufficient learning of ID embeddings in lifelong sequence and multi-modal alignment. For online fast approximate retrieval, a multi-modal product-quantization based strategy is also proposed. Extensive evaluations on multiple datasets demonstrate the effectiveness of our method.

\bibliographystyle{ACM-Reference-Format}
\bibliography{final_paper_v1}


\begin{thebibliography}{22}


\ifx \showCODEN    \undefined \def \showCODEN     #1{\unskip}     \fi
\ifx \showDOI      \undefined \def \showDOI       #1{#1}\fi
\ifx \showISBNx    \undefined \def \showISBNx     #1{\unskip}     \fi
\ifx \showISBNxiii \undefined \def \showISBNxiii  #1{\unskip}     \fi
\ifx \showISSN     \undefined \def \showISSN      #1{\unskip}     \fi
\ifx \showLCCN     \undefined \def \showLCCN      #1{\unskip}     \fi
\ifx \shownote     \undefined \def \shownote      #1{#1}          \fi
\ifx \showarticletitle \undefined \def \showarticletitle #1{#1}   \fi
\ifx \showURL      \undefined \def \showURL       {\relax}        \fi
\providecommand\bibfield[2]{#2}
\providecommand\bibinfo[2]{#2}
\providecommand\natexlab[1]{#1}
\providecommand\showeprint[2][]{arXiv:#2}

\bibitem[\protect\citeauthoryear{Ai, Zhang, Bi, Chen, and Croft}{Ai et~al\mbox{.}}{2017}]%
        {10.1145/3077136.3080813}
\bibfield{author}{\bibinfo{person}{Qingyao Ai}, \bibinfo{person}{Yongfeng Zhang}, \bibinfo{person}{Keping Bi}, \bibinfo{person}{Xu Chen}, {and} \bibinfo{person}{W.~Bruce Croft}.} \bibinfo{year}{2017}\natexlab{}.
\newblock \showarticletitle{Learning a Hierarchical Embedding Model for Personalized Product Search}. In \bibinfo{booktitle}{\emph{Proceedings of the 40th International ACM SIGIR Conference on Research and Development in Information Retrieval}} (Shinjuku, Tokyo, Japan) \emph{(\bibinfo{series}{SIGIR '17})}. \bibinfo{publisher}{Association for Computing Machinery}, \bibinfo{address}{New York, NY, USA}, \bibinfo{pages}{645–654}.
\newblock
\showISBNx{9781450350228}
\urldef\tempurl%
\url{https://doi.org/10.1145/3077136.3080813}
\showDOI{\tempurl}


\bibitem[\protect\citeauthoryear{Chang, Zhang, Fu, Zang, Guan, Lu, Hui, Leng, Niu, Song, and Gai}{Chang et~al\mbox{.}}{2023}]%
        {10.1145/3580305.3599922}
\bibfield{author}{\bibinfo{person}{Jianxin Chang}, \bibinfo{person}{Chenbin Zhang}, \bibinfo{person}{Zhiyi Fu}, \bibinfo{person}{Xiaoxue Zang}, \bibinfo{person}{Lin Guan}, \bibinfo{person}{Jing Lu}, \bibinfo{person}{Yiqun Hui}, \bibinfo{person}{Dewei Leng}, \bibinfo{person}{Yanan Niu}, \bibinfo{person}{Yang Song}, {and} \bibinfo{person}{Kun Gai}.} \bibinfo{year}{2023}\natexlab{}.
\newblock \showarticletitle{TWIN: TWo-stage Interest Network for Lifelong User Behavior Modeling in CTR Prediction at Kuaishou}. In \bibinfo{booktitle}{\emph{Proceedings of the 29th ACM SIGKDD Conference on Knowledge Discovery and Data Mining}} (<conf-loc>, <city>Long Beach</city>, <state>CA</state>, <country>USA</country>, </conf-loc>) \emph{(\bibinfo{series}{KDD '23})}. \bibinfo{publisher}{Association for Computing Machinery}, \bibinfo{address}{New York, NY, USA}, \bibinfo{pages}{3785–3794}.
\newblock
\showISBNx{9798400701030}
\urldef\tempurl%
\url{https://doi.org/10.1145/3580305.3599922}
\showDOI{\tempurl}


\bibitem[\protect\citeauthoryear{Chen, Xu, Pei, Lv, Zhuang, and Ge}{Chen et~al\mbox{.}}{2022}]%
        {chen2022efficient}
\bibfield{author}{\bibinfo{person}{Qiwei Chen}, \bibinfo{person}{Yue Xu}, \bibinfo{person}{Changhua Pei}, \bibinfo{person}{Shanshan Lv}, \bibinfo{person}{Tao Zhuang}, {and} \bibinfo{person}{Junfeng Ge}.} \bibinfo{year}{2022}\natexlab{}.
\newblock \bibinfo{title}{Efficient Long Sequential User Data Modeling for Click-Through Rate Prediction}.
\newblock
\newblock
\showeprint[arxiv]{2209.12212}~[cs.IR]


\bibitem[\protect\citeauthoryear{Devlin, Chang, Lee, and Toutanova}{Devlin et~al\mbox{.}}{2019}]%
        {devlin2019bert}
\bibfield{author}{\bibinfo{person}{Jacob Devlin}, \bibinfo{person}{Ming-Wei Chang}, \bibinfo{person}{Kenton Lee}, {and} \bibinfo{person}{Kristina Toutanova}.} \bibinfo{year}{2019}\natexlab{}.
\newblock \bibinfo{title}{BERT: Pre-training of Deep Bidirectional Transformers for Language Understanding}.
\newblock
\newblock
\showeprint[arxiv]{1810.04805}~[cs.CL]


\bibitem[\protect\citeauthoryear{Dosovitskiy, Beyer, Kolesnikov, Weissenborn, Zhai, Unterthiner, Dehghani, Minderer, Heigold, Gelly, Uszkoreit, and Houlsby}{Dosovitskiy et~al\mbox{.}}{2020}]%
        {DBLP:journals/corr/abs-2010-11929}
\bibfield{author}{\bibinfo{person}{Alexey Dosovitskiy}, \bibinfo{person}{Lucas Beyer}, \bibinfo{person}{Alexander Kolesnikov}, \bibinfo{person}{Dirk Weissenborn}, \bibinfo{person}{Xiaohua Zhai}, \bibinfo{person}{Thomas Unterthiner}, \bibinfo{person}{Mostafa Dehghani}, \bibinfo{person}{Matthias Minderer}, \bibinfo{person}{Georg Heigold}, \bibinfo{person}{Sylvain Gelly}, \bibinfo{person}{Jakob Uszkoreit}, {and} \bibinfo{person}{Neil Houlsby}.} \bibinfo{year}{2020}\natexlab{}.
\newblock \showarticletitle{An Image is Worth 16x16 Words: Transformers for Image Recognition at Scale}.
\newblock \bibinfo{journal}{\emph{CoRR}}  \bibinfo{volume}{abs/2010.11929} (\bibinfo{year}{2020}).
\newblock
\showeprint[arXiv]{2010.11929}
\urldef\tempurl%
\url{https://arxiv.org/abs/2010.11929}
\showURL{%
\tempurl}


\bibitem[\protect\citeauthoryear{Douze, Guzhva, Deng, Johnson, Szilvasy, Mazaré, Lomeli, Hosseini, and Jégou}{Douze et~al\mbox{.}}{2024}]%
        {douze2024faiss}
\bibfield{author}{\bibinfo{person}{Matthijs Douze}, \bibinfo{person}{Alexandr Guzhva}, \bibinfo{person}{Chengqi Deng}, \bibinfo{person}{Jeff Johnson}, \bibinfo{person}{Gergely Szilvasy}, \bibinfo{person}{Pierre-Emmanuel Mazaré}, \bibinfo{person}{Maria Lomeli}, \bibinfo{person}{Lucas Hosseini}, {and} \bibinfo{person}{Hervé Jégou}.} \bibinfo{year}{2024}\natexlab{}.
\newblock \showarticletitle{The Faiss library}.
\newblock  (\bibinfo{year}{2024}).
\newblock
\showeprint[arxiv]{2401.08281}~[cs.LG]


\bibitem[\protect\citeauthoryear{Gong, Ding, Su, Shen, Liu, and Zhang}{Gong et~al\mbox{.}}{2023}]%
        {10.1145/3583780.3614657}
\bibfield{author}{\bibinfo{person}{Yuqi Gong}, \bibinfo{person}{Xichen Ding}, \bibinfo{person}{Yehui Su}, \bibinfo{person}{Kaiming Shen}, \bibinfo{person}{Zhongyi Liu}, {and} \bibinfo{person}{Guannan Zhang}.} \bibinfo{year}{2023}\natexlab{}.
\newblock \showarticletitle{An Unified Search and Recommendation Foundation Model for Cold-Start Scenario}. In \bibinfo{booktitle}{\emph{Proceedings of the 32nd ACM International Conference on Information and Knowledge Management}} (<conf-loc>, <city>Birmingham</city>, <country>United Kingdom</country>, </conf-loc>) \emph{(\bibinfo{series}{CIKM '23})}. \bibinfo{publisher}{Association for Computing Machinery}, \bibinfo{address}{New York, NY, USA}, \bibinfo{pages}{4595–4601}.
\newblock
\showISBNx{9798400701245}
\urldef\tempurl%
\url{https://doi.org/10.1145/3583780.3614657}
\showDOI{\tempurl}


\bibitem[\protect\citeauthoryear{Guo, Li, Yang, Liang, Yuan, Hou, Ke, Zhang, He, Zhang, Yu, and Ou}{Guo et~al\mbox{.}}{2023}]%
        {10.1145/3583780.3615022}
\bibfield{author}{\bibinfo{person}{Tong Guo}, \bibinfo{person}{Xuanping Li}, \bibinfo{person}{Haitao Yang}, \bibinfo{person}{Xiao Liang}, \bibinfo{person}{Yong Yuan}, \bibinfo{person}{Jingyou Hou}, \bibinfo{person}{Bingqing Ke}, \bibinfo{person}{Chao Zhang}, \bibinfo{person}{Junlin He}, \bibinfo{person}{Shunyu Zhang}, \bibinfo{person}{Enyun Yu}, {and} \bibinfo{person}{Wenwu Ou}.} \bibinfo{year}{2023}\natexlab{}.
\newblock \showarticletitle{Query-dominant User Interest Network for Large-Scale Search Ranking}. In \bibinfo{booktitle}{\emph{Proceedings of the 32nd ACM International Conference on Information and Knowledge Management}} (<conf-loc>, <city>Birmingham</city>, <country>United Kingdom</country>, </conf-loc>) \emph{(\bibinfo{series}{CIKM '23})}. \bibinfo{publisher}{Association for Computing Machinery}, \bibinfo{address}{New York, NY, USA}, \bibinfo{pages}{629–638}.
\newblock
\showISBNx{9798400701245}
\urldef\tempurl%
\url{https://doi.org/10.1145/3583780.3615022}
\showDOI{\tempurl}


\bibitem[\protect\citeauthoryear{Hou, He, McAuley, and Zhao}{Hou et~al\mbox{.}}{2023}]%
        {hou2023learning}
\bibfield{author}{\bibinfo{person}{Yupeng Hou}, \bibinfo{person}{Zhankui He}, \bibinfo{person}{Julian McAuley}, {and} \bibinfo{person}{Wayne~Xin Zhao}.} \bibinfo{year}{2023}\natexlab{}.
\newblock \bibinfo{title}{Learning Vector-Quantized Item Representation for Transferable Sequential Recommenders}.
\newblock
\newblock
\showeprint[arxiv]{2210.12316}~[cs.IR]


\bibitem[\protect\citeauthoryear{Jégou, Douze, and Schmid}{Jégou et~al\mbox{.}}{2011}]%
        {5432202}
\bibfield{author}{\bibinfo{person}{Herve Jégou}, \bibinfo{person}{Matthijs Douze}, {and} \bibinfo{person}{Cordelia Schmid}.} \bibinfo{year}{2011}\natexlab{}.
\newblock \showarticletitle{Product Quantization for Nearest Neighbor Search}.
\newblock \bibinfo{journal}{\emph{IEEE Transactions on Pattern Analysis and Machine Intelligence}} \bibinfo{volume}{33}, \bibinfo{number}{1} (\bibinfo{year}{2011}), \bibinfo{pages}{117--128}.
\newblock
\urldef\tempurl%
\url{https://doi.org/10.1109/TPAMI.2010.57}
\showDOI{\tempurl}


\bibitem[\protect\citeauthoryear{Malkov and Yashunin}{Malkov and Yashunin}{2018}]%
        {malkov2018efficient}
\bibfield{author}{\bibinfo{person}{Yu.~A. Malkov} {and} \bibinfo{person}{D.~A. Yashunin}.} \bibinfo{year}{2018}\natexlab{}.
\newblock \bibinfo{title}{Efficient and robust approximate nearest neighbor search using Hierarchical Navigable Small World graphs}.
\newblock
\newblock
\showeprint[arxiv]{1603.09320}~[cs.DS]


\bibitem[\protect\citeauthoryear{Pi, Zhou, Zhang, Wang, Ren, Fan, Zhu, and Gai}{Pi et~al\mbox{.}}{2020}]%
        {10.1145/3340531.3412744}
\bibfield{author}{\bibinfo{person}{Qi Pi}, \bibinfo{person}{Guorui Zhou}, \bibinfo{person}{Yujing Zhang}, \bibinfo{person}{Zhe Wang}, \bibinfo{person}{Lejian Ren}, \bibinfo{person}{Ying Fan}, \bibinfo{person}{Xiaoqiang Zhu}, {and} \bibinfo{person}{Kun Gai}.} \bibinfo{year}{2020}\natexlab{}.
\newblock \showarticletitle{Search-based User Interest Modeling with Lifelong Sequential Behavior Data for Click-Through Rate Prediction}. In \bibinfo{booktitle}{\emph{Proceedings of the 29th ACM International Conference on Information \& Knowledge Management}} (Virtual Event, Ireland) \emph{(\bibinfo{series}{CIKM '20})}. \bibinfo{publisher}{Association for Computing Machinery}, \bibinfo{address}{New York, NY, USA}, \bibinfo{pages}{2685–2692}.
\newblock
\showISBNx{9781450368599}
\urldef\tempurl%
\url{https://doi.org/10.1145/3340531.3412744}
\showDOI{\tempurl}


\bibitem[\protect\citeauthoryear{Radford, Kim, Hallacy, Ramesh, Goh, Agarwal, Sastry, Askell, Mishkin, Clark, Krueger, and Sutskever}{Radford et~al\mbox{.}}{2021}]%
        {radford2021learning}
\bibfield{author}{\bibinfo{person}{Alec Radford}, \bibinfo{person}{Jong~Wook Kim}, \bibinfo{person}{Chris Hallacy}, \bibinfo{person}{Aditya Ramesh}, \bibinfo{person}{Gabriel Goh}, \bibinfo{person}{Sandhini Agarwal}, \bibinfo{person}{Girish Sastry}, \bibinfo{person}{Amanda Askell}, \bibinfo{person}{Pamela Mishkin}, \bibinfo{person}{Jack Clark}, \bibinfo{person}{Gretchen Krueger}, {and} \bibinfo{person}{Ilya Sutskever}.} \bibinfo{year}{2021}\natexlab{}.
\newblock \bibinfo{title}{Learning Transferable Visual Models From Natural Language Supervision}.
\newblock
\newblock
\showeprint[arxiv]{2103.00020}~[cs.CV]


\bibitem[\protect\citeauthoryear{Rajput, Mehta, Singh, Keshavan, Vu, Heldt, Hong, Tay, Tran, Samost, Kula, Chi, and Sathiamoorthy}{Rajput et~al\mbox{.}}{2023}]%
        {rajput2023recommender}
\bibfield{author}{\bibinfo{person}{Shashank Rajput}, \bibinfo{person}{Nikhil Mehta}, \bibinfo{person}{Anima Singh}, \bibinfo{person}{Raghunandan~H. Keshavan}, \bibinfo{person}{Trung Vu}, \bibinfo{person}{Lukasz Heldt}, \bibinfo{person}{Lichan Hong}, \bibinfo{person}{Yi Tay}, \bibinfo{person}{Vinh~Q. Tran}, \bibinfo{person}{Jonah Samost}, \bibinfo{person}{Maciej Kula}, \bibinfo{person}{Ed~H. Chi}, {and} \bibinfo{person}{Maheswaran Sathiamoorthy}.} \bibinfo{year}{2023}\natexlab{}.
\newblock \bibinfo{title}{Recommender Systems with Generative Retrieval}.
\newblock
\newblock
\showeprint[arxiv]{2305.05065}~[cs.IR]


\bibitem[\protect\citeauthoryear{Si, Sun, Zhang, Xu, Zang, Song, Gai, and Wen}{Si et~al\mbox{.}}{2023}]%
        {10.1145/3539618.3591786}
\bibfield{author}{\bibinfo{person}{Zihua Si}, \bibinfo{person}{Zhongxiang Sun}, \bibinfo{person}{Xiao Zhang}, \bibinfo{person}{Jun Xu}, \bibinfo{person}{Xiaoxue Zang}, \bibinfo{person}{Yang Song}, \bibinfo{person}{Kun Gai}, {and} \bibinfo{person}{Ji-Rong Wen}.} \bibinfo{year}{2023}\natexlab{}.
\newblock \showarticletitle{When Search Meets Recommendation: Learning Disentangled Search Representation for Recommendation}. In \bibinfo{booktitle}{\emph{Proceedings of the 46th International ACM SIGIR Conference on Research and Development in Information Retrieval}} (, Taipei, Taiwan,) \emph{(\bibinfo{series}{SIGIR '23})}. \bibinfo{publisher}{Association for Computing Machinery}, \bibinfo{address}{New York, NY, USA}, \bibinfo{pages}{1313–1323}.
\newblock
\showISBNx{9781450394086}
\urldef\tempurl%
\url{https://doi.org/10.1145/3539618.3591786}
\showDOI{\tempurl}


\bibitem[\protect\citeauthoryear{Sun, Si, Zang, Leng, Niu, Song, Zhang, and Xu}{Sun et~al\mbox{.}}{2023}]%
        {kuaisar_dataset}
\bibfield{author}{\bibinfo{person}{Zhongxiang Sun}, \bibinfo{person}{Zihua Si}, \bibinfo{person}{Xiaoxue Zang}, \bibinfo{person}{Dewei Leng}, \bibinfo{person}{Yanan Niu}, \bibinfo{person}{Yang Song}, \bibinfo{person}{Xiao Zhang}, {and} \bibinfo{person}{Jun Xu}.} \bibinfo{year}{2023}\natexlab{}.
\newblock \showarticletitle{KuaiSAR: A Unified Search And Recommendation Dataset}. In \bibinfo{booktitle}{\emph{Proceedings of the 32nd ACM International Conference on Information and Knowledge Management}} (<conf-loc>, <city>Birmingham</city>, <country>United Kingdom</country>, </conf-loc>) \emph{(\bibinfo{series}{CIKM '23})}. \bibinfo{publisher}{Association for Computing Machinery}, \bibinfo{address}{New York, NY, USA}, \bibinfo{pages}{5407–5411}.
\newblock
\showISBNx{9798400701245}
\urldef\tempurl%
\url{https://doi.org/10.1145/3583780.3615123}
\showDOI{\tempurl}


\bibitem[\protect\citeauthoryear{van~den Oord, Vinyals, and Kavukcuoglu}{van~den Oord et~al\mbox{.}}{2017}]%
        {10.5555/3295222.3295378}
\bibfield{author}{\bibinfo{person}{Aaron van~den Oord}, \bibinfo{person}{Oriol Vinyals}, {and} \bibinfo{person}{Koray Kavukcuoglu}.} \bibinfo{year}{2017}\natexlab{}.
\newblock \showarticletitle{Neural discrete representation learning}. In \bibinfo{booktitle}{\emph{Proceedings of the 31st International Conference on Neural Information Processing Systems}} (Long Beach, California, USA) \emph{(\bibinfo{series}{NIPS'17})}. \bibinfo{publisher}{Curran Associates Inc.}, \bibinfo{address}{Red Hook, NY, USA}, \bibinfo{pages}{6309–6318}.
\newblock
\showISBNx{9781510860964}


\bibitem[\protect\citeauthoryear{Vaswani, Shazeer, Parmar, Uszkoreit, Jones, Gomez, Kaiser, and Polosukhin}{Vaswani et~al\mbox{.}}{2023}]%
        {vaswani2023attention}
\bibfield{author}{\bibinfo{person}{Ashish Vaswani}, \bibinfo{person}{Noam Shazeer}, \bibinfo{person}{Niki Parmar}, \bibinfo{person}{Jakob Uszkoreit}, \bibinfo{person}{Llion Jones}, \bibinfo{person}{Aidan~N. Gomez}, \bibinfo{person}{Lukasz Kaiser}, {and} \bibinfo{person}{Illia Polosukhin}.} \bibinfo{year}{2023}\natexlab{}.
\newblock \bibinfo{title}{Attention Is All You Need}.
\newblock
\newblock
\showeprint[arxiv]{1706.03762}~[cs.CL]


\bibitem[\protect\citeauthoryear{Yao, Dou, Xie, Lu, Wang, and Wen}{Yao et~al\mbox{.}}{2021}]%
        {10.1145/3459637.3482489}
\bibfield{author}{\bibinfo{person}{Jing Yao}, \bibinfo{person}{Zhicheng Dou}, \bibinfo{person}{Ruobing Xie}, \bibinfo{person}{Yanxiong Lu}, \bibinfo{person}{Zhiping Wang}, {and} \bibinfo{person}{Ji-Rong Wen}.} \bibinfo{year}{2021}\natexlab{}.
\newblock \showarticletitle{USER: A Unified Information Search and Recommendation Model based on Integrated Behavior Sequence}. In \bibinfo{booktitle}{\emph{Proceedings of the 30th ACM International Conference on Information \& Knowledge Management}} (Virtual Event, Queensland, Australia) \emph{(\bibinfo{series}{CIKM '21})}. \bibinfo{publisher}{Association for Computing Machinery}, \bibinfo{address}{New York, NY, USA}, \bibinfo{pages}{2373–2382}.
\newblock
\showISBNx{9781450384469}
\urldef\tempurl%
\url{https://doi.org/10.1145/3459637.3482489}
\showDOI{\tempurl}


\bibitem[\protect\citeauthoryear{Zeghidour, Luebs, Omran, Skoglund, and Tagliasacchi}{Zeghidour et~al\mbox{.}}{2021}]%
        {zeghidour2021soundstream}
\bibfield{author}{\bibinfo{person}{Neil Zeghidour}, \bibinfo{person}{Alejandro Luebs}, \bibinfo{person}{Ahmed Omran}, \bibinfo{person}{Jan Skoglund}, {and} \bibinfo{person}{Marco Tagliasacchi}.} \bibinfo{year}{2021}\natexlab{}.
\newblock \bibinfo{title}{SoundStream: An End-to-End Neural Audio Codec}.
\newblock
\newblock
\showeprint[arxiv]{2107.03312}~[cs.SD]


\bibitem[\protect\citeauthoryear{Zhao}{Zhao}{2023}]%
        {10.1145/3580305.3599863}
\bibfield{author}{\bibinfo{person}{Pengyu et~al. Zhao}.} \bibinfo{year}{2023}\natexlab{}.
\newblock \showarticletitle{M5: Multi-Modal Multi-Interest Multi-Scenario Matching for Over-the-Top Recommendation}. In \bibinfo{booktitle}{\emph{Proceedings of the 29th ACM SIGKDD Conference on Knowledge Discovery and Data Mining}} (<conf-loc>, <city>Long Beach</city>, <state>CA</state>, <country>USA</country>, </conf-loc>) \emph{(\bibinfo{series}{KDD '23})}. \bibinfo{publisher}{Association for Computing Machinery}, \bibinfo{address}{New York, NY, USA}, \bibinfo{pages}{3785–3794}.
\newblock
\showISBNx{9798400701030}
\urldef\tempurl%
\url{https://doi.org/10.1145/3580305.3599863}
\showDOI{\tempurl}


\bibitem[\protect\citeauthoryear{Zhou, Wang, Zhao, Zhu, Wang, Zhang, Wang, and Wen}{Zhou et~al\mbox{.}}{2020}]%
        {S_3_rec_ssl}
\bibfield{author}{\bibinfo{person}{Kun Zhou}, \bibinfo{person}{Hui Wang}, \bibinfo{person}{Wayne~Xin Zhao}, \bibinfo{person}{Yutao Zhu}, \bibinfo{person}{Sirui Wang}, \bibinfo{person}{Fuzheng Zhang}, \bibinfo{person}{Zhongyuan Wang}, {and} \bibinfo{person}{Ji-Rong Wen}.} \bibinfo{year}{2020}\natexlab{}.
\newblock \showarticletitle{S3-Rec: Self-Supervised Learning for Sequential Recommendation with Mutual Information Maximization}. In \bibinfo{booktitle}{\emph{Proceedings of the 29th ACM International Conference on Information \& Knowledge Management}} (Virtual Event, Ireland) \emph{(\bibinfo{series}{CIKM '20})}. \bibinfo{publisher}{Association for Computing Machinery}, \bibinfo{address}{New York, NY, USA}, \bibinfo{pages}{1893–1902}.
\newblock
\showISBNx{9781450368599}
\urldef\tempurl%
\url{https://doi.org/10.1145/3340531.3411954}
\showDOI{\tempurl}


\end{thebibliography}

\end{document}